\documentclass[12pt,a4paper]{article}

\usepackage{latexsym}
\usepackage{graphicx,color}
\usepackage{amsmath}
\newcommand\sect[1]{\section{#1}\setcounter{equation}0}

\newcommand\no{\nonumber\\}
\newcommand\eqnb{\begin{eqnarray}}
\newcommand\eqne{\end{eqnarray}}
\newcommand\ad{{\rm ad}}

\newcommand\onematrix{\mbox{\footnotesize $1\!\!$}1}
\usepackage{latexsym}

\begin{document}
\center{{\huge Stretched quantum membranes}\vspace{15mm}\\
Jonas Bj\" ornsson \footnote{jonas.bjornsson@kau.se} and Stephen Hwang 
\footnote{stephen.hwang@kau.se}\\Department of Physics\\Karlstad University
\\SE-651 88 Karlstad, Sweden}\vspace{15mm}\\

\abstract{We study bosonic and space-time supersymmetric membranes with small tensions corresponding to 
stretched configurations. Using a generalized lightcone gauge, one may set up a perturbation theory 
around configurations having 
zero tension. We will show, by explicit construction to all orders in perturbation theory, 
that these membrane configurations 
are canonically equivalent, and thereby solvable, to string-like configurations with 
string excitations transverse to the stretched
direction.  
At the quantum level, it is shown that there exists an ordering such that 
equivalence by unitary transformations is achieved. Consistency requires
the critical dimensions 27 and 11 for the bosonic and supersymmetric cases, respectively. 
The mass spectrum is determined to any order. It is discrete and contains massless exitations. 
The ground state is purely string-like, whereas excited string-like states 
split through the perturbation into an infinite set of states with equal or lower energies.}
\newpage


\sect{Introduction}
The relativistic membrane is, unlike string theory at the first quantized level, a highly complex theory.
It is self-interacting and solutions to the equations of motion are, therefore, hard to find. Still,
it is an interesting theory for many reasons. It is a theory in the same family as string theory, i.e.\ a geometrical
theory where the action is proportional to the world-volume. As such it is interesting to study whether or not
lessons learnt from string theory generalize, or if there are new and interesting features that are uncovered.

The membrane is also interesting from the point of M-theory \cite{Townsend:1995kk,Witten:1995ex}. 
In \cite{Townsend:1995kk} M-theory was conjectured to be the supermembrane. A further connection between the two
was established
through the matrix model based on the work of \cite{goldstone,hoppe,deWit:1988ig} and the conjecture in 
\cite{Banks:1996vh}. In the former, a discretization of the area-preserving 
diffeomorphism algebra of the supermembrane action in lightcone gauge, gives a maximally 
supersymmetric $SU(N)$ matrix model. This 
matrix model should then in the large-$N$ limit recover the full dynamics of the supermembrane. 
The connection to the same matrix model proposed in \cite{Banks:1996vh}, using the discrete lightcone approach, 
was further discussed in e.g.\ \cite{Seiberg:1997} and
\cite{Sen:98}. 
 
In view of the interest in the membrane and the complexity of the theory, any new results that may shed new light
to its properties are highly valuable. In a previous paper \cite{Bjornsson:2004yp} we formulated a 
perturbative approach
to studying membranes around infinitely stretched configurations with zero tension. The perturbation parameter is 
the membrane tension.
The unperturbed theory behaves as a string theory with string-like\footnote{It is equivalent to 
the string theory apart from the fact that it depends on a third world-volume parameter. We use, therefore 
the terminology "string-like".} excitations transverse to the stretching.
Consequently, one may exactly solve the unperturbed theory. String-like configurations have
also been found in the context of D-branes at strong coupling \cite{Lindstrom:1997uj}.

In \cite{Bjornsson:2004yp} we proposed to solve the perturbation theory by canonical transformations, transforming 
the perturbed theory into the unperturbed one. In the present paper, we will prove that this proposal is indeed possible
to implement to any order in perturbation theory. We will also treat the supermembrane, generalizing the 
result to this case as well. This implies that these membrane configurations, with small tensions, are classically
equivalent to string-like theories and, therefore, completely solvable in terms of string-like solutions. Our results
hold for open, semi-open and closed membranes.

Having shown the canonical equivalence at the classical level makes it possible to address the problem of quantization. We will show that 
it is possible to define a specific ordering such that the canonical transformations become unitary ones. Then
the quantum theory requires, to any order in perturbation theory, the critical dimensions
$D=27$ and $D=11$ for the bosonic membrane and supermembrane, respectively. 

Evidence 
for these critical dimensions are indirectly present through the double-dimensional reduction \cite{Duff:1987bx},
was further investigated
for the supermembrane at the massless \cite{Bars:1987dy} and first massive level \cite{Bars:1987nr}, and 
discussed in connection with the BRST symmetry in \cite{Fujikawa:1987av}.
In \cite{Marquard:1988bj,Marquard:1989rd} these critical dimensions were found using Weyl-ordering and 
point-splitting regularization. In our case the critical dimensions arise from 
requirement that the $(D-1)$-dimensional Lorentz symmetry is 
non-anomalous. Indirectly, it is an implication that a two-dimensional subalgebra of the three-parameter 
reparametrization symmetry is non-anomalous for the critical dimensions to any order in 
perturbation theory. It is
still an open question if our construction yields membranes that have full 
Lorentz invariance and reparametrization invariance at the quantum level.

We can construct, using the unitary transformations and to any order in perturbation theory,
an infinite set of physical states that diagonalizes the Hamiltonian. We will determine the spectrum 
to any order in perturbation theory. The spectrum is discrete and contains massless excitations. 
 This is the case for both the bosonic and the supersymmetric membrane. 
The unperturbed spectrum is the usual string spectrum where each excited level is infinitely degenerate. 
The perturbation will break the degeneracy creating, in addition to the string-like states, an infinite set of states with lower mass. 
In particular, we will show that the mass splitting is the same for any order. 

To arrive at this result, we have implicitly assumed a certain class of boundary conditions. The 
spectrum 
will depend crucially on this. In particular, for the supermembrane our choice 
seems to be the only one yielding a massless groundstate.
Furthermore, it cannot be ruled out that for other boundary conditions it may be possible to have 
a continous spectrum. 

Our results for the stretched membrane may be compared to results found in the matrix model 
approximation. For
example, it has been shown that the bosonic membrane has a discrete spectrum \cite{Simon,deWit:1988ct}.
For the supermembrane one has, in general, a continous spectrum \cite{deWit:1988ct}. However, in 
\cite{Aref'eva:1999ku} 
it was shown that for the $SU(2)$ supersymmetric matrix model there exists different possibilities. One 
possibility is a purely discrete spectrum and another is a spectrum which has a continous as well as a 
discrete part. The continous mass spectrum for the supermembrane can be understood in terms of
the existence of infinitely thin tubes or stretches in the membrane surface \cite{deWit:1988ct}. For 
the bosonic case there is an infinite potential barrier preventing such effects. It has been 
proposed that the pinching of surfaces implies that supermembranes are second quantized from the outset. 

In our approach we will see that there is a barrier against pinching.
The stretched membranes will always have a minimum width or circumference. This is a consequence of our particular 
gauge choice, keeping the 
string-like tension of the unperturbed theory fixed. Our results are consistent with the matrix 
model calculations
for the bosonic case. For the supersymmetric case our results disagree with the matrix model 
calculations discussed above. This is true even for the case when the matrix model 
gives a
discrete spectrum, since then there are no massless states \cite{Aref'eva:1999ku}. 
The disagreement could mean that our assumption regarding boundary conditions are not 
compatible with matrix model ones. 
It could also imply that the matrix model calculations probe the theory outside the weak tension regime, 
in which our perturbation theory is defined. Further investigations are certainly needed to answer 
these questions.

It is clear that in our perturbative approach we do not see any interactions between the 
string-like excitations. Interactions could arise as
non-perturbative effects in our treatment. In a generic membrane configuration one may find stretched
configurations in parts of the membrane surface eg.\ in thin tubes connecting different sections of 
the membrane, or in spikes that are attached to the surface. In these parts of the surface the excitations
would, according to our results here, be purely string-like. An attractive interpretation of the string-like
excitations would be as the elementary excitations of the membrane. In particular,
the spikes could be thought of as in- and outgoing asymptotic string-like states. 
The non-stretched parts of the 
membrane surface correspond to sections of the surface in which interactions could take place.

The paper is organized as follows. In section two we make a short review of our formulation in 
\cite{Bjornsson:2004yp} yielding a perturbation theory around infinitely stretched configurations.
In section three we fix the gauge completely by
using a generalized lightcone gauge and show the canonical equivalence to all 
orders in perturbation theory. We generalize our result to the supermembrane in section four. 
In section five we use the results in the previous sections to define a 
quantum theory which is unitarily equivalent to a string-like theory. The paper is concluded in the 
last section. In appendix A we give a few conventions and in appendix B we present some calculations
on a toy model, that captures some of the relevant features of the membrane.


\sect{The basic formalism and previous results}

The Dirac action \cite{Dirac} of the membrane is
\begin{equation}
S=-T_{m}\int d^3\xi\left[-\det\left(\partial_i X_U\partial_j X^U\right)\right]^{1/2},
\label{action}
\end{equation}
where we use the mostly plus convension of the metric, $U=0,\ldots, D-1$ and $i,j=0,1,2$. $\xi^i$ 
parametrize the world volume where $\xi^0$ is the time-like parameter. 
This action has three constraints, corresponding to reparametrization invariance of 
the world-volume. 

By choosing a partial gauge $\chi\equiv X^{D-2}-\frac{1}{\sqrt{g}} \xi^2 \approx 0$, where $g$ is a constant, the Hamiltonian may 
be put into the form \cite{Bjornsson:2004yp}
\eqnb
H&=&\int d^2\xi \phi_2\no
&=&H_0+gH_1,\label{hamiltonian0}
\eqne
with first-class constraints
\begin{eqnarray}
\phi_1&=&\mathcal P \partial_1 X\approx 0
\label{phi1 constraint}\\
\phi_2&=&\frac{1}{2}\left\{\mathcal P^2+\left(\partial_1 X\right)^2\right.\no
&+&
\left.
g\left[
\left(\partial_1 X\right)^2\left(\partial_2 X\right)^2+\left(\mathcal P 
\partial_2 X\right)^2-\left(\partial_1 X \partial_2 X\right)^2\right]\right\} \approx 0.
\label{phi2 constraint}
\end{eqnarray}
All scalar products are now in the $D-1$ dimensional space-time $\mu=0,\ldots , D-3,D-1$. We have, as can be seen from eqs. 
(\ref{hamiltonian0}), (\ref{phi1 constraint}) and (\ref{phi2 constraint}) for $g\ll 1$, 
a perturbation theory, where the unperturbed
$g=0$ limit yields a string-like theory with string-like tension\footnote{The string-like tension has 
dimension, $(length)^{-3}$, which is different from the ordinary string tension.}, $T_s=1$. In arriving at this 
formulation, we have taken 
$g=\left(\frac{T_m}{T_s}\right)^2$ so that $T_m=\sqrt{g}\ll 1=T_s$. Thus, $g\rightarrow 0$ is a zero-tension limit
of the membrane such that the string tension in the $\xi^1$-direction remains non-zero. As the gauge choice 
$X^{D-2}-\frac{1}{\sqrt{g}} \xi^2 \approx 0$ implies that
$X^{D-2}\rightarrow \infty$ when $g\rightarrow 0$, we see that the membrane fills this space-direction completely.
Thus, it is stretched in this direction. If the membrane is closed in the $\xi^2$-direction, the gauge-choice is
consistent for $X^{D-2}$ being periodic. We will mostly consider
a membrane that is periodic in both parameters i.e. have a torus topology. Other possibilites may be 
treated analogously and the results for these cases are indicated.

In \cite{Bjornsson:2004yp} we studied the Hamiltonian defined in 
eq.\ (\ref{hamiltonian0}).
We were able to show the existence of canonical transformations that transformed the perturbed Hamiltonian 
to the unperturbed one to any order in perturbation theory. Since the gauge was only partially
fixed, the effect of the transformations on the remaining constraints was not considered. 

The canonical transformations in \cite{Bjornsson:2004yp} also produced additional terms that involved the fields evaluated 
at the boundary. These terms are potentially problematic, as they contain factors that are non-local in the
the space-direction corresponding to the boundary and, hence, may 
contribute to the equations of motion away from the boundary. This was not properly realized in \cite{Bjornsson:2004yp}.
It can be seen that, to first order in perturbation
theory, the explicit expression in \cite{Bjornsson:2004yp} does not give such problems, as the boundary terms actually
vanish by the constraints. This was an implication that, by modifying the transformation, 
it should be possible to avoid
generating boundary terms. The situation at arbitrary order was, however, an open question. 
The transformations constructed here will not generate boundary terms for the open case.


\sect{The Bosonic membrane}
We will in this section show that one can map the perturbed theory to the unperturbed, string-like, theory by a canonical 
transformation. We do this by completely fixing a gauge for the bosonic membrane action. 

Let us begin with the constraints for the membrane action
\eqnb
\phi_0&=&\frac{1}{2}\mathcal P^2+\frac{T_m^2}{2}\left[\left(\partial_1 X\right)^2
\left(\partial_2 X\right)^2-
\left(\partial_1X \cdot \partial_2 X\right)^2\right]\approx 0\\
\phi_1&=&\mathcal P \partial_1 X\approx 0\no
\phi_2&=&\mathcal P\partial_2 X\approx 0.
\eqne
Fix a gauge partially by using the lightcone gauge
\eqnb
\chi_1&\equiv&X^+-\xi^0\approx 0
\no
\chi_2&\equiv&\mathcal P^+-1\approx 0,
\label{lightcone}
\eqne
where we have used the conventions in appendix A. From this procedure 
one can define a Hamiltonian from $\mathcal P^-$
\eqnb
H&=&-\int d^2\xi\mathcal P^-\no
&=&\frac12\int d^2\xi\left\{\mathcal P^2+T_m\left[(\partial_1 X)^2(\partial_2 X)^2-
(\partial_1 X\cdot\partial_2 X)^2\right]\right\},
\eqne
where the scalar products are in the $D-2$ space-like dimensions. One  constraint remains ungauged
\eqnb
\phi&=&\partial_1 \mathcal P\cdot\partial_2 X-\partial_2 \mathcal P\cdot\partial_1 X\approx 0.
\eqne
To gauge fix this, and to define our perturbation theory, we choose
\eqnb
\chi_3&\equiv&X^{D-2}-\frac{1}{\sqrt g}\xi^2\approx 0.
\label{our gauge}
\eqne
$X^{D-2}$ has to be compact if one considers a closed membrane in 
this direction.
Choosing $g=T_m^2$ this gauge defines a 
perturbation theory for $g\ll 1$ around a string-like theory in the lightcone gauge
\eqnb
H&=&H_0+gH_1\no
H_0&=& \frac12 \int d^2\xi\left[\mathcal P^2+\partial_1 X^2\right]\no
H_1&=&\frac{1}{2}\int d^2\xi\left[(\partial_1 X)^2(\partial_2 X)^2
+B^2-(\partial_1 X\partial_2 X)^2\right]\label{hamiltonians}
\eqne
where $B\equiv -\frac{1}{\sqrt g}\mathcal P^{D-2}$ with
\eqnb
\partial_ 1B&=&\partial_1 \mathcal P\cdot\partial_2 X-\partial_2 \mathcal P\cdot\partial_1 X
\label{B}
\eqne
and all scalar products are now in $D-3$ dimensions. These equations will be the starting point for our
analysis of the perturbation theory.

Let us make a change of basis by
\eqnb
\alpha^I&=&\frac{1}{\sqrt 2}\left(\mathcal P^I+\partial_1 X^I\right)\no
\tilde\alpha^I&=&\frac{1}{\sqrt 2}\left(\mathcal P^I-\partial_1 X^I\right).
\eqne
Using the fundamental Poisson bracket
\eqnb
\left\{X^I(\xi),\mathcal P^I(\xi')\right\}&=&\delta^{IJ}\delta^{2}\left(\xi-\xi'\right),
\eqne
one has
\eqnb
\left\{\alpha^I(\xi),\alpha^J(\xi')\right\}&=&
\frac{1}{2}\delta^{IJ}\left(\partial_1-\partial'_1\right)\delta^{2}(\xi-\xi')\no
\left\{\tilde\alpha^I(\xi),\tilde\alpha^J(\xi')\right\}&=&
-\frac{1}{2}\delta^{IJ}\left(\partial_1-\partial'_1\right)\delta^{2}(\xi-\xi').
\eqne
We make a Fourier expansion of the $\xi^1$-dependence of the fields $\alpha^I$ and $\tilde\alpha^I$
\eqnb
\alpha^I&=&\frac{1}{\sqrt{2\pi}}\sum_{m}\alpha^I_m\exp[-im\xi^1]\no
\tilde\alpha^I&=&\frac{1}{\sqrt{2\pi}}\sum_{m}\tilde\alpha^I_m\exp[im\xi^1],\label{fourier}
\eqne
where $\alpha^I_m$ and $\tilde\alpha^I_m$ depend on $\xi^0$ and $\xi^2$. 
The non-zero Poisson brackets of the Fourier 
coefficients are 
\eqnb
\left\{\alpha^I_m(\xi^2),\alpha^J_n(\xi'^2)\right\}&=&-im\delta_{m,-n}\delta^{IJ}\delta(\xi^2-\xi'^2)\no
\left\{\tilde\alpha^I_m(\xi^2),\tilde\alpha^J_n(\xi'^2)\right\}&=&-im\delta_{m,-n}\delta^{IJ}\delta(\xi^2-\xi'^2).
\eqne
Using eq.\ (\ref{fourier}) we have the Fourier expansions
\eqnb
\mathcal P^I&=&\frac{1}{2\sqrt\pi}\sum_{n}\left(\alpha_n^I\exp[-in\xi^1]+\tilde\alpha^I_n\exp[in\xi^1]\right)
\label{P_n}
\\
\partial_1 X^I&=&\frac{1}{2\sqrt\pi}\sum_{n}\left(\alpha_n^I\exp[-in\xi^1]-\tilde\alpha^I_n\exp[in\xi^1]\right)
\eqne
Integrating the last equation, requiring $X^I$ to be periodic, yields
\eqnb
X^I&=&q^I
+
\frac{i}{2\sqrt\pi}\sum_{n\neq 0}\frac{1}{n}\left(\alpha_n^I\exp[-in\xi^1]-\tilde\alpha^I_n\exp[in\xi^1]\right),
\label{X_n}
\eqne
$\alpha^I_0=\tilde\alpha^I_0$. From the fundamental Poisson brackets we have
\eqnb
\left\{q^I(\xi^2),\alpha_0^J({\xi'}^2)\right\}&=&\frac{1}{2\sqrt\pi}\delta^{IJ}\delta(\xi^2-\xi'^2).
\eqne
The unperturbed Hamiltonian is of the string-like form
\eqnb
H_0&=&\frac{1}{2}\sum_{m,I}\int d\xi^2\left(\alpha_{-m}^I\alpha_{m}^I+\tilde\alpha_{-m}^I\tilde\alpha_{m}^I\right).
\eqne
We may also determine $B$,
\eqnb
B&=&\frac{1}{2\sqrt\pi}\sum_{m,I}\left(\partial_2q^I\alpha_m^I\exp\left[-im\xi^1\right]+
\partial_2q^I\tilde\alpha_m^I\exp\left[im\xi^1\right]\right)\no
&+&
\frac{i}{4\pi}\sum_{n\neq 0,m,I}\frac{1}{n}\left(\alpha^I_m\partial_2\alpha^I_n\exp\left[-i(m+n)\xi^1\right]
+\tilde\alpha^I_m\partial_2\alpha^I_n\exp\left[i(m-n)\xi^1\right]\right.\no
&+&
\left.
\alpha^I_m\partial_2\tilde\alpha^I_n\exp\left[-i(m-n)\xi^1\right]
+\tilde\alpha^I_m\partial_2\tilde\alpha^I_n\exp\left[i(m+n)\xi^1\right]\right)\no
&-&\frac{i}{2\pi}\sum_{m+n\neq 0,I}\frac{1}{m+n}\left(
\alpha_m^I\partial_2\alpha_n^I\exp\left[-i(n+m)\xi^1\right]
\right.
\no
&+&
\left.\tilde\alpha_m^I\partial_2\tilde\alpha_n^I\exp\left[i(n+m)\xi^1\right]\right),
\label{B field}
\eqne
up to a function $f(\xi^2)$. This function will depend on the fourier modes of our basic 
fields through boundary conditions. We will take this function to be zero. Consistency will 
restrict the possible choices of boundary conditions. Other choices will not change our results apart from 
the spectrum. At the end of section five, we will briefly discuss the case of a non-zero 
function and possible implications on the mass spectrum.

If one now studies the perodicity 
condition on $X^-$ one finds
\eqnb
\sum_{m,I}\left[\alpha^I_{-m}\alpha^I_{m}-\tilde\alpha^I_{-m}\tilde\alpha^I_{m}\right]&=&0,
\label{LR}
\eqne
which is also a sufficient condition for $B$ to be periodic. Using $B$ one can explicitely 
find the expression for $H_1$. 

Let us now consider the problem that we should solve. We would like to find a canonical
transformation that transforms the perturbed theory at an arbitrary order to the unperturbed one.
As $g\ll 1$ it is sufficient to consider infinitesimal transformations. To first order in perturbation
theory we should, therefore, find a generator $G_1$ of infinitesimal canonical transformations that satisfies
\eqnb
H\rightarrow H'=H+g\{H,G_1\}=H_0+\mathcal O(g^2)
\eqne
which implies
\eqnb
\{H_0,G_1\}=-H_1.\label{1-eq}
\eqne
As can be seen from the explicit expression of $H_1$ in eq.\ (\ref{hamiltonians}) it is composed of 
quartic terms in the Fourier modes. $H_0$ is quadratic so that $G_1$ will also be quartic. To 
next order in perturbation theory one should find a generator $G_2$ such that 
\eqnb
H\rightarrow H'=H+g\{H,G_1\}+\frac{g^2}{2}\{\{H,G_1\},G_1\}+g^2\{H,G_2\}=H_0+\mathcal O(g^3)
\eqne
which implies
\eqnb
\{H_0,G_2\}=-\{H_1,G_1\}-\frac{1}{2}\{\{H_0,G_1\},G_1\}\equiv -H_2.
\eqne
$G_2$ will essentially be a sum of terms of products of six Fourier modes. Continuing this order by order, the $N$'th 
order generator will satisfy an equation of the form
\eqnb
\{H_0,G_N\}=-H_N.\label{N-eq}
\eqne
$G_N$ will essentially be a sum of terms of products of $2(N+1)$ Fourier modes. Even though $H_N$ becomes more and more
complicated as $N$ grows larger, it is still possible to prove that a solution exists. The reason is that 
$G_N$ is determined through
its Poisson bracket with $H_0$, which has a very simple form. 

Beginning with the first order, $H_1$ may in a compact way be written as
\eqnb
H_1=\int d\xi^2\sum_{r}\left\{\mathcal H^0_{(r)}
+\partial_2q_{I} \mathcal H^{I}_{(r)}
+\partial_2q_{I}\partial_2 q_{J}
\mathcal H^{IJ}_{(r)}\right\},
\label{ham-1}
\eqne
$\mathcal H^0_{(r)}$, $\mathcal H^{I}_{(r)}$ and 
$\mathcal H^{IJ}_{(r)}$ are sums of terms with products of
four, three and two factors
of $\alpha_m^I$, $\tilde\alpha_m^I$ and derivatives with respect to $\xi^2$ of these fields
satisfying
\eqnb
\left\{H_0,\mathcal H^{ (\ldots )}_{(r)}\right\}&=& ir \mathcal H^{(\ldots )}_{(r)}.
\label{calH}
\eqne

We first consider the problem of getting rid of a term $\mathcal H^0_{(r)}$ in $H_1$.
For $r$ different from zero it is simple to find a solution and it is 
\eqnb
G_1^{part.}&=&-\frac{1}{ir}\int d\xi^2\mathcal H^0_{(r)}.
\eqne
Consider now the case when $r=0$. Introduce a quantity $K$ satisfying
\eqnb
\left\{H_0,K\right\}&=&-1.\label{H-K}
\eqne
Such a quantity exists. Let us construct it by first calculating the Poisson bracket between 
$\int d\xi^2 k_Iq^I$, where $k^I$ is a constant vector, and $H_0$
\eqnb
\left\{\int d\xi^2\alpha_{I,0}(\xi^2)\alpha^I_0(\xi^2),\int d\xi'^2 k_J(\xi')q^J(\xi')\right\}
=-\frac{1}{\sqrt\pi}\int d\xi^2k_I\alpha^I_0.
\eqne
Thus, $K$ is given by
\eqnb
K&\equiv &\sqrt\pi\frac{\int d\xi^2 k_Iq^I}{\int d\xi^2k_I\alpha^I_0},\label{K}
\eqne
which is well-defined if the centre of mass momentum 
$P^I_0= \int d\xi^2\alpha^I_0$ is non-zero. 
As $P^I_0$ is a constant of motion this requirement restricts the initial state of the membrane. 
One can now use this operator to solve the terms 
where $r=0$ by
\eqnb
G_1^{part.}&=&K\int d\xi^2\mathcal H^0_{(0)}.\label{g1k}
\eqne

Let us now focus on more complicated terms containing $q^I$'s in $H_1$. Consider the term in
$H_1$ containing one $q^I$. Let us make an ansatz of the solution for $r\neq 0$ as
\eqnb
\tilde G_1^{part.}&=&-\frac{1}{ir}\int d\xi^2\partial_2 q_I\mathcal H_{(r)}^I.
\eqne
The Poisson bracket of this term with $H_0$ gives a term that is the sought for term. 
In addition, it gives a term
\eqnb
-\frac{1}{ir}\int d\xi^2\left(-\frac{\partial_2\alpha_{I,0}}{\sqrt\pi}\right)\mathcal H_{(r)}^I.
\eqne
This term is of the same kind as the one considered above, which we have shown how to compensate 
by a canonical transformation. We can, therefore, write down the solution to the term with one $q^I$ 
and $r\neq 0$,
\eqnb
G_1^{part.}=-\frac{1}{ir}\int d\xi^2\partial_2 q_I\mathcal H_{(r)}^I+
\frac{1}{(ir)^2}\int d\xi^2\left(-\frac{\partial_2 \alpha_{I,0}}{\sqrt\pi}\right)\mathcal H^I_{(r)}.
\eqne
When $r$ is equal to zero one can proceed in the same way. 
The solution for this part of $G_1$ is
\eqnb
G_1^{part.}&=&K\int d\xi^2\partial_2q_I\mathcal H_{(0)}^I+
\frac{K^2}{2}\int d\xi^2\left(-\frac{\partial_2\alpha_{I,0}}{\sqrt\pi}\right)\mathcal H_{(0)}^I.
\eqne
The third term in eq.\ (\ref{ham-1}), containing two factors of $q$, is solved 
in a completely analogous fashion. Let us conclude by presenting the solution to first order
\eqnb
G_1&=&\int d\xi^2\sum_{r}\left\{\left(-\frac{1}{ir}\left(1-\delta_{r,0}\right)+
K\delta_{r,0}\right)\left(\mathcal H^{0}_{(r)}
+\partial_2 q_I \mathcal H^I_{(r)}\right.\right.
\no
&+&\left.\partial_2 q_I\partial_2 q_J \mathcal H^{IJ}_{(r)}\right)
+\left(\frac{1}{(ir)^2}\left(1-\delta_{r,0}\right)+\frac{K^2}{2}\delta_{r,0}\right)
\no
&\times&
\left( \left(-\frac{\partial_2 \alpha_{I,0}}{\sqrt\pi}\right) \mathcal H^I_{(r)}
+2\left(-\frac{\partial_2 \alpha_{I,0}}{\sqrt\pi}\right)\partial_2 q_J \mathcal H^{IJ}_{(r)}
\right)\no
&+&
2\left(-\frac{1}{(ir)^3}\left(1-\delta_{r,0}\right)+\frac{K^3}{6}\delta_{r,0}\right)
\left.
\left(-\frac{\partial_2 \alpha_{I,0}}{\sqrt\pi}\right)
\left(-\frac{\partial_2 \alpha_{J,0}}{\sqrt\pi}\right) \mathcal H^{IJ}_{(r)}\right\}.\no
\label{canonical fourier solution}
\eqne
We have by this expression constructed the solution of the generator of canonical transformation to 
first order. 
In the remaining part of this section we will show how one can generalize the solution (\ref{canonical 
fourier solution}) to get an expression to any order in 
perturbation theory. It is convenient to make a Fourier decomposition 
with respect to the $\xi^2$-dependence,
\eqnb
\alpha^I_m&=&\frac{1}{\sqrt{2\pi}}\sum_{n} \alpha^I_{m,n}\exp\left[in\xi^2\right]\no
\tilde\alpha^I_m&=&\frac{1}{\sqrt{2\pi}}\sum_{n} \tilde\alpha^I_{m,n}
\exp\left[in\xi^2\right]\no
q^I&=&\frac{1}{{2}^{3/2}\pi}\sum_{n}q^I_n\exp\left[in\xi^2\right].
\eqne
We have the following non-zero Poisson brackets
\eqnb
\left\{\alpha_{m,n}^I,\alpha_{p,q}^J\right\}&=&-im\delta_{m,-p}\delta_{n,-q}\delta^{IJ}\no
\left\{\tilde\alpha_{m,n}^I,\tilde\alpha_{p,q}^J\right\}&=&-im\delta_{m,-p}\delta_{n,-q}\delta^{IJ}\no
\left\{q_m^I,\alpha_{0,n}^J\right\}&=&\delta_{m,-n}\delta^{IJ}.
\eqne
From this one gets that the quantity $K$ has the form
\eqnb
K&=&\frac{1}{2}\frac{k_I q^I_0}{k_I\alpha^I_{0,0}}\label{KF}
\eqne
To simplify the expressions, let us define a collective index $(a)=(I,n)$
\eqnb
\alpha^I_{m,n}&=&\alpha^{(a)}_m
\eqne
In this notation we have the unperturbed Hamiltonian
\eqnb
H_0&=&\frac{1}{2}\sum_{m,(a)}\left(\alpha^{(-a)}_{m}\alpha^{(a)}_{m}
+\tilde\alpha^{(-a)}_{m}\tilde\alpha^{(a)}_{m}\right),
\eqne
where $(-a)=(I,-n)$ for $(a)=(I,n)$, and the first order Hamiltonian (cf.\ equation (\ref{ham-1}))
\eqnb
H_1&=&\sum_{r}\left\{H^0_{(r)}+\sum_{(a)}q^{(-a)}H^{(a)}_{(r)}+
\sum_{(a),(b)}q^{(-a)}q^{(-b)}H^{(a)(b)}_{(r)}\right\}.
\label{H1 fourier}
\eqne

Consider the following general form of the Hamiltonian to $N$'th order
\eqnb
H_{N}&=&\sum_{r}\sum_{i=0}\sum_{(a_1)\ldots (a_i)}
\left\{q^{(-a_1)}\cdot\ldots\cdot q^{(-a_i)}H^{(a_1)\ldots(a_i)}_{(r)}\right\}.
\label{HN}
\eqne
$H^{(a_1)\ldots(a_i)}_{(r)}\in \mathcal F$, which is 
the space of polynomials of $\alpha^{(a_p)}_{m_p}$, $\tilde\alpha_{m_p}^{(a_p)}$
and $\left(k_I\alpha_{0,0}^I\right)^{-1}$. Define also $\mathcal G$ as the space of 
polynomials of $\alpha^{(a_p)}_{m_p}$, $\tilde\alpha_{m_p}^{(a_p)}$, $\left(k_I\alpha_{0,0}^I\right)^{-1}$ 
and $q^I$, so 
that $H_{N}\in \mathcal G$.
We will prove below that this is the most general form of $H_N$ that can appear.
It is true to first order, which can be seen from eq.\ (\ref{H1 fourier}). Before we proceed, let us 
note the following 
\eqnb
\left\{q^{(a)},f\right\}&\in& \mathcal F\;{\rm{for}}\;f\in\mathcal F \no
\left\{f,f'\right\}&\in& \mathcal F\;{\rm{for}}\;f,f'\in\mathcal F.
\eqne
This implies 
\eqnb
\left\{H^{\{1\}}_{N_1},H^{\{2\}}_{N_2}\right\}&\in & \mathcal G,
\label{HS}
\eqne
where $H^{\{i\}}_{N_i}\in \mathcal G$.
We also need to construct the canonical transformation which yields the Hamiltonian eq.\ (\ref{HN}). 
Proceeding in the same way as we did to first order, it is straightforward to construct the solution. 
One finds
\eqnb
G_N&=&\sum_{r}G^{r}_N,
\label{GN}
\eqne
where
\eqnb
G^{r\neq 0}_N &=&
\sum_{i=0}\sum_{j=0}^{i}\sum_{(a_1)\ldots(a_i)}
\binom{p}{j}
\frac{(-1)^{j+1}}{(ir)^{j+1}}
\left\{\left(-2\alpha_0^{(-a_1)}\right)\cdot\ldots\right.\no
&\times&
\left.\left(-2\alpha_0^{(-a_j)}\right)
q^{(-a_{j+1})}\cdot\ldots\cdot q^{(-a_i)}H^{(a_1)\ldots(a_i)}_{(r)}
\right\}
\label{GN{r neq 0}}
\\
G^{r=0}_N&=&
\sum_{i=0}\sum_{j=0}^{i}\sum_{(a_1)\ldots(a_i)}
\binom{p}{j}
\frac{K^j}{j!}
\left\{\left(-2\alpha_0^{(-a_1)}\right)\cdot\ldots\cdot\left(-2\alpha_0^{(-a_j)}\right)\right.\no
&\times&
\left.q^{(-a_{j+1})}\cdot\ldots\cdot q^{(-a_i)}
H^{(a_1)\ldots(a_i)}_{(0)}
\right\}.
\label{GN{r = 0}}
\eqne
By inspection one sees that $G_N\in 
\mathcal G$. 
Therefore, by eq.\ (\ref{HS})
\eqnb
\left\{G^{\{1\}}_{N_1},H^{\{2\}}_{N_2}\right\}&\in & \mathcal G,
\label{GS}
\eqne
where $G^{\{1\}}_{N_1}\in  \mathcal G$ and $H^{\{2\}}_{N_2}\in \mathcal G$.

Let us now prove that eq.\ (\ref{HN}) is the most general form of the Hamiltonian. The proof is by induction.
It is obviously true for the zeroth and first orders. $H_{N+1}$ is constructed by taking repeated 
Poisson brackets between generators $G_i$, where $i=1,\ldots, N$, and $H_0$. By the induction hypothesis 
$G_i\in \mathcal G$. Using equation (\ref{GS}) repeatedly proves the statement.

We have now shown that, to any order in perturbation theory, the Hamiltonian has the form given by 
eq.\ (\ref{HN}) 
and constructed the generators of canonical transformations, given by eqs. (\ref{GN}), 
(\ref{GN{r neq 0}}) and (\ref{GN{r = 0}}), which transforms $H$ to $H_0$. We have here 
considered the closed case. Extending the analysis to the open case is straightforward. By an 
appropriate boundary condition, we will get a Fourier expansion w.r.t. $\xi^1$ with only one 
independent mode, $\alpha_n^I$. The proof we give above holds also for this case.
In the next section we will extend our analysis to also hold for the completely gauge fixed 
supermembrane.

 
\sect{The Supermembrane}
We will in this section generalize our results to the space-time supersymmetric membrane. 
Consider the following action proposed  in  \cite{Bergshoeff:1987cm} 
\eqnb
S&=&-T_m\int d^3\xi\left\{\sqrt{-h}+
i\frac 12 \epsilon^{ijk}\bar \theta \Gamma_{UV}\partial_i \theta\right.\no
&\times&\left.
\left[\Pi^U_j\partial_k X^V-\frac 13 \bar\theta\Gamma^U\partial_j
\theta\bar\theta\Gamma^V\partial_k\theta\right]\right\}.
\eqne
We specify the dimension to $D=11$, generalization to $D=4,5$ and $7$ is straightforward at the classical level. 
Here $X^U$ are 
the $11$ bosonic coordinates,  $\theta^\beta$  are 32-component Majorana 
spinors,  
$\Pi_i^U=\partial_iX^U-i\bar\theta\Gamma^U\partial_i\theta$, $h$ is the 
determinant of the matrix $h_{ij}=\Pi_i^U\Pi_{j,U}$, $\Gamma^U$ are  
gamma matrices, $\Gamma_{UV}=\frac12\left[\Gamma_U,\Gamma_V\right]$ 
and $\epsilon^{012}=-1$. This action is invariant under local fermionic transformations 
(kappa symmetry) and local reparametrizations. Let us first fix the kappa symmetry by
\eqnb
\Gamma^+\theta&=&0
\label{projection},
\eqne
where we use the lightcone conventions presented in the appendix A. 
Using the explicit basis of $\Gamma^{+}$, given in Appendix A, one can see that $\theta^{2\beta+1}=0$. We also rescale 
the other fermions by $\theta^{2\beta}=2^{-1/4}\psi^\beta$. Passing on to the phase-space 
formulation, using the lightcone gauge, 
defined in eq.\ (\ref{lightcone}), yields the Hamiltonian \cite{deWit:1988ig}
\eqnb
H&=&\int d^2\xi\left\{\frac{1}{2}\mathcal P^2+
T_m i\epsilon^{ab}\psi\gamma_A\partial_a\psi\partial_bX^A
\right.
\no
&+&
\left.\frac{T^2_{m}}{2}\left[\left(\partial_1 X\right)^2\left(\partial_2 X\right)^2
-\left(\partial_1 X\cdot\partial_2 X\right)^2\right]\right\}.
\eqne
We also have the remaining constraints
\eqnb
\phi&=&\epsilon^{ab}\left(\partial_a\mathcal P^A\partial_bX_A+
i\partial_a \psi\partial_b \psi\right)\approx 0\no
G^\beta&=&S^\beta-i\psi^\beta\approx 0,
\eqne
where $G^\beta$ are all second-class. Eliminating the second-class constraints gives rise 
to a non-zero Dirac bracket between the fermions
\eqnb
\left\{\psi^\beta(\xi),\psi^\gamma(\xi')\right\}^\ast&=&
\frac{1}{2i}\delta^{\beta\gamma}\delta^{2}(\xi-\xi').
\eqne
We eliminate the bosonic constraint by imposing the gauge in eq.\ (\ref{our gauge}). Denote also 
$g=T^2_{m}$. Letting $g$ be small yields a perturbation theory where the Hamiltonian 
can be divided into three parts as
\eqnb
H&=&H_0+\sqrt g H_1+gH_2\no
H_0&=&\int d^2\xi\left\{\frac12\left[\mathcal P^2+\left(\partial_1 X\right)^2\right]
+i\psi_1\partial_1\psi_1-i\psi_2\partial_1\psi_2\right\}\no
H_1&=&\int d^2\xi i\epsilon^{ab}\psi\gamma_I\partial_a\psi\partial_bX^I\no
H_2&=&\frac{1}{2}\int d^2\xi\left[\left(\partial_1 X\right)^2\left(\partial_2 X\right)^2+\tilde B^2
-\frac 12 \left(\partial_1 X \partial_2 X\right)^2\right],
\eqne
where $\tilde B\equiv -\frac{1}{\sqrt{g}}\mathcal P^{D-2}$ and
\eqnb
\psi&=&
\left(
\begin{array}{c}
\psi^1\\
\psi^2
\end{array}
\right),
\eqne
with
\eqnb
\partial_1 \tilde B&=&-\frac{1}{\sqrt g}\partial_1 \mathcal P^{D-2}\no
&=&
\partial_1\mathcal P\cdot\partial_2 X+i\partial_1\psi_1 \partial_2 \psi_1 -
\partial_2\mathcal P\cdot\partial_1 X-i\partial_2\psi_2 \partial_1 \psi_2.
\label{tilde B}
\eqne
Note here the 
difference as compared to the bosonic membrane. Here we get order $\sqrt g$ 
corrections, which are terms with one $\xi^2$-derivative. 

Let us proceed in the same manner as in the previous section. Change coordinates from 
$(\partial_1 X^I,P^I)$ to $(\alpha^I,\tilde\alpha^I)$. Expanding these into Fourier modes yields in the 
end eqs. (\ref{P_n}) and (\ref{X_n}) with $\alpha_0^I=\tilde \alpha_0^I$. Similarly, for the fermions
\eqnb
\psi^1_\beta&=&\frac{1}{2\sqrt{\pi}}\sum_{n}\psi^1_{\beta,n}\exp[-in\xi^1]\no
\psi^2_\beta&=&\frac{1}{2\sqrt{\pi}}\sum_{n}\psi^2_{\beta,n}\exp[in\xi^1],
\eqne
where $\psi^1_{\beta,n}$ and $\psi^2_{\beta,n}$ depend on $\xi^0$ and $\xi^2$. The non-zero Poisson 
brackets are
\eqnb
\left\{\psi^1_{\beta,m},\psi^1_{\gamma,n}\right\}
&=&-i\delta_{m,-n}\delta_{\beta\gamma}\delta(\xi^2-\xi'^2)\no
\left\{\psi^2_{\beta,m},\psi^2_{\gamma,n}\right\}
&=&-i\delta_{m,-n}\delta_{\beta\gamma}\delta(\xi^2-\xi'^2)
\eqne
Inserting this into the unperturbed Hamiltonian yields
\eqnb
H_0=\frac{1}{2}\sum_{m}\int d^2\xi\left[\left(\alpha_{-m}^I\alpha_{I,m}+\alpha_{-m}^I\alpha_{I,m}\right)
+m\left(\psi^1_{-m}\psi^1_{m}+\psi^2_{-m}\psi^2_{m}\right)\right].
\eqne
Before we proceed, let us determine the explicit expression of $\tilde B$, which we calculate by 
inserting the Fourier expansions of the different fields
\eqnb
\tilde B&=&B+\frac{i}{4\pi}\sum_{n,m}\left\{\psi^1_m\partial_2\psi^1_n\exp\left[-i(m+n)\xi^1\right]
+\psi^2_m\partial_2\psi^2_n\exp\left[i(m+n)\xi^1\right]
\right\}\no
&-&
\frac{i}{2\pi}\sum_{n+m\neq 0}\frac{n}{n+m}\left\{\psi^1_m\partial_2\psi^1_n\exp\left[-i(m+n)\xi^1\right]
\right.\no
&+&
\left.
\psi^2_m\partial_2\psi^2_n\exp\left[i(m+n)\xi^1\right]\right\},
\label{tilde B field}
\eqne
where $B$ is given by eq.\ (\ref{B field}). The periodicity requires
\eqnb
\sum_{m}\left[\alpha^I_{-m}\alpha_{I,m}+m\psi^1_{-m}\psi^1_{m}
-\tilde\alpha^I_{-m}\tilde\alpha_{I,m}-m\psi^2_{-m}\psi^2_{m}\right]&=&0.
\eqne
If we study the Poisson brackets between the unperturbed Hamiltonian and terms of the form
\eqnb
\mathcal H_{(r)}\left(\xi^2\right)&=&C_{I_1,\ldots,I_N}^{\alpha_{N+1},\ldots,\alpha_R}(n_i)\alpha^{I_1}_{n_1}\cdot\ldots\cdot\alpha^{I_M}_{n_M}
\cdot\tilde\alpha^{I_{M+1}}_{n_{M+1}}\cdot\ldots\cdot\tilde\alpha^{I_N}_{n_N}\no
&\times&\psi^1_{n_{N+1},\alpha_{N+1}}
\cdot\ldots\cdot\psi^1_{n_{P},\alpha_P}\psi^2_{n_{P+1},
\alpha_{P+1}}\cdot\ldots\cdot\psi^2_{n_R,\alpha_R},\no
r&=&\sum_{i=1}^R{n_i},
\eqne
one finds
\eqnb
\left\{H_0,\mathcal H_{(r)}\right\}&=&ir\mathcal H_{(r)}.
\eqne
This also holds true for terms that involve derivatives with respect to $\xi^2$. 
Therefore, $\mathcal H_{(r)}$ 
satisfies the same crucial Poisson bracket as the corresponding expression for the bosonic 
case, eq.\ (\ref{calH}). We can then
make a decomposition of different terms that can appear in $H_N$ in the same way as in 
eq.\ (\ref{HN}) where, in addition, $\mathcal F$ and $\mathcal G$ contains fermion modes $\psi^i_{\beta,n}$. 
Furthermore, the quantity $K$ defined in eq.\ (\ref{K}) still satisfies the property eq.\ (\ref{H-K}). 
This means that the construction of $G_N$ proceeds in exactly the same way as for the bosonic case 
with a general solution given by eqs. (\ref{GN}), (\ref{GN{r neq 0}}) and (\ref{GN{r = 0}}). 
This generalizes also to the open case.


\sect{Quantization}

Our considerations this far have been purely classical. In this section we will 
show that there exist, as far as we have been able to check, consistent quantum theories to any finite order
in perturbation theory for stretched membranes in $D=27$ and $D=11$ for the bosonic and supersymmetric
cases, respectively. 

The preceeding sections show that the membrane is, in a particular gauge
and within our perturbation theory, canonically equivalent to a string-like theory. This does 
not automatically mean that there is a quantum equivalence, due to ordering
problems. Every canonical transformation will not correspond to a unitary transformation. 
Our approach is, however, to turn the argument around and begin by considering a consistent 
quantum theory, namely the string-like theory, and letting this theory define an ordering that makes it possible to 
define a consistent
quantum membrane through unitary transformations.

Our starting point, therefore, will be to consider the $g=0$ theory where we simply have a string-like theory.
Consider for simplicity the bosonic case. The supersymmetric case is completely analogous. Choose also,
for definitness, a closed membrane in the $\xi^1$-direction. The open case is treated in the same way.
Solving the equations of motion, with periodic boundary conditions, one has 
quantum mechanically the string-like commutator for the left-moving modes
\eqnb
[\alpha_m^I(\xi^2), {\alpha}_n^J({\xi'}^2)]=m\delta_{m,-n}\delta^{IJ}\delta (\xi^2-{\xi'}^2),
\eqne
and a corresponding one for the right-moving sector. The zeroth order groundstate of the membrane 
is given by 
the string-like groundstate $\alpha_m^I(\xi^2)\mid 0,p_0\rangle_{0} =p^I_0\delta_{m,0}\mid 0,p_0\rangle_{0}$ and 
$\tilde{\alpha}_m^I(\xi^2)\mid 0,p_0\rangle_{0}=p^I_0\delta_{m,0}\mid 0,p_0\rangle_{0}$, for 
$p^I_0$ being a constant and $m\geq 0$. This implies that 
$\alpha_{m,n}^I\mid 0,p_0\rangle_{0} =p^I_0\delta_{m,0}\delta_{n,0}\mid 0,p_0\rangle_{0}$, for $m\geq 0$ and all 
$n$, etc., if we Fourier expand w.r.t. the 
$\xi^2$-coordinate. 
Next, let us construct a unitary operator, $U_1$, by
\eqnb
U_1\equiv \exp(-ig:_0 G_{1}:_0).
\eqne
Here $G_1$ is the generator constructed in section three, classically transforming away terms of 
order one of the 
membrane Hamiltonian, leaving the zeroth order string-like one. The notation $:_0$ indicates a normal 
ordering w.r.t. 
the zeroth order groundstate i.e.\ the string-like one. We also order so that $q^I_n$ is on the left of 
$\alpha^I_{0,n}$, $n\neq 0$, and $q^I_0$ and 
$\alpha^I_{0,0}$ are Weyl ordered.
In particular, $K$ defined in eq.\ (\ref{KF}), should be defined as an operator as 
\eqnb
K&=&\frac{1}{4}\left[{k_I q^I_0}\left(k_I\alpha^I_{0,0}\right)^{-1}+\left(k_I\alpha^I_{0,0}\right)^{-1}{k_I q^I_0}
\right].
\eqne
We now define a Hamiltonian 
\eqnb
H_{membrane}^{(1)}\equiv U_1 :_0H_{string}:_0U^\dagger_1.
\eqne
In the classical limit, $H_{membrane}^{(1)}$ will become the membrane Hamiltonian in eq.\ (\ref{hamiltonian0}) to order one
in perturbation theory. Define the first order groundstate 
\eqnb
\left|0,p_0\right>_1&=&U_1\left|0,p_0\right>_{0}.
\eqne
It satisfies $\alpha_{m,n}^{(1),I}\mid 0,p_0\rangle_{1} =p^I_0\delta_{m,0}\delta_{n,0}\mid 0,p_0\rangle_{1}$, 
for $m\geq 0$ and all $n$, where $\alpha_{m,n}^{(1),I}\equiv U_1\alpha_{m,n}^{I}U^\dagger_1$.

Proceeding to second order, we take a unitary transformation 
\eqnb
U_2\equiv \exp(-ig^2:_1G_{2}:_1),
\eqne
where $:_1$ is normal ordering w.r.t.\ the first order groundstate. Define a Hamiltonian 
\eqnb
H_{membrane}^{(2)}=U_2U_1 :_0H_{string}:_0U^\dagger_1U^\dagger_2.
\eqne
$H_{membrane}^{(2)}$ is classically equivalent to the membrane Hamiltonian to order two in perturbation
theory. Using this procedure iteratively, one will arrive at a Hamiltonian that, in the classical 
limit, gives the membrane Hamiltonian to any order in perturbation theory. This construction of the 
membrane Hamiltonian gives a specific ordering of the operators, that from a membrane point of 
view is non-trivial. In Appendix B a toy model with similar properties is treated. The treatment gives
the explicit construction up to second order.

The groundstate that one gets to order $N$ is
\eqnb
\mid 0,p_0\rangle_{N}=U_N\ldots U_1\mid 0,p_0\rangle_{0}, \label{vac}
\eqne
which satisfies that 
\eqnb
\alpha^{(N),I}_{m,n}\mid 0,p_0\rangle_{N} =p_0^I\delta_{m,0}\delta_{n,0}\mid 0,p_0\rangle_{N},
\eqne
for $m\geq 0$ and all $n$, where
\eqnb
\alpha^{(N),I}_{m,n}=
U_N\ldots U_1 \alpha^{I}_{m,n} U^\dagger_N\ldots U^\dagger_1.
\eqne
Note that 
$:_0H_{string}:_0\mid 0, p_0\rangle _N,$
is, in general, infinite for $N\geq 1$. This implies that the 
perturbation in $g$ is non-perturbative from a string point of view.
Note also that $K$ is a well-defined operator in eq.\ (\ref{vac}) for $p^I_0\neq 0$. 
Eigenstates to the Hamiltonian may also be constructed in the same fashion. Let
\eqnb
\mid \phi\rangle_{0}=\hat \phi\mid 0,p_0\rangle_{0}
\eqne
be any string-like eigenstate. Then, obviously,
\eqnb
\mid \phi\rangle_{N}=U_N\ldots U_1\hat\phi\mid 0,p_0\rangle_{0}
\eqne
is an eigenstate to the Hamiltonian to order $N$ in perturbation theory. 
$\mid \phi\rangle_{0}$ consists of 
all states given by applying the creation operators $\alpha^I_{-m,n}$, $\tilde\alpha^I_{-m,n}$ 
and $\exp(i\sum_n \tilde k_{-n,I}q^I_{n})$, for $m>0$, all $n$ and constant $\tilde k_{n,I}$, 
to the groundstate such that eq.\ (\ref{LR}) is satisfied.

Our theory is obviously invariant under the $(D-3)$-dimensional Lorentz group. But, in addition, 
one has $(D-1)$-dimensional 
Lorentz symmetry provided $D-1=26$ or $D-1=10$ i.e.\ $D=27$ and $D=11$, respectively. 
This follows trivially from the fact that the unperturbed string-like
theory has this unbroken symmetry for the critical dimensions and the unitary equivalence of stretched
membranes to string-like configurations at any order in perturbation theory. 

This does not, however, prove the existence of a consistent quantum membrane theory with 27- or 11-dimensional 
Lorentz invariance. 
The commutators that need to be checked for the closure of full Lorentz algebra are those which 
involve at least one of $M^{D-2,\,I}$, $M^{D-2,\,+}$ and $M^{D-2,\,-}$, $I=1,\ldots 24$ or 
$I=1,\ldots 8$. 
These generators are, for the bosonic case and at the classical level, of the form 
\eqnb
M^{D-2,\,I}&=&\frac{1}{\sqrt{g}}\left[\int d^2\xi P^I\xi^2 +g\int d^2\xi X^I B\right]\no
M^{D-2,\,+}&=&\frac{1}{\sqrt{g}}\left[2\pi^2+g\xi^0\int d^2\xi B\right]\no
M^{D-2,\,-}&=&\frac{1}{\sqrt{g}}\left[-\int d^2\xi \mathcal H\xi^2+g\int d^2\xi X_0^- B \right],
\eqne
where $H=\int d^2\xi \mathcal H$, $P_0^U=\int d^2\xi\mathcal P^U$ etc. $X^-$ is solved from
the constraints and contains terms of zeroth order as well as first order terms in $g$. 
If one 
computes the commutators, at the quantum level and up to order $\mathcal O(g^{-1/2})$, it is 
straightforward to see that they
are anomaly free. To next order, terms of order $g^{0}$, the calculation is complicated 
since different orders will be mixed. Furthermore, the unitary transformations
will not simplify the calculations.
The quantum consistency of the full Lorentz algebra is, therefore, 
still an open and non-trivial question.

Let us now discuss the mass spectrum that is implied by our construction. The unitary 
transformations, that transform
the membrane states into string-like states, ensure that, from a $(D-1)$-dimensional point of view, 
the mass spectrum will be exactly the
same as for string theory. There will, however, be an infinite degeneracy corresponding to the 
$\xi^2$-dependence. From
a $D$-dimensional point of view the mass spectrum is slightly different.
The mass is determined from the operator 
\eqnb
m^2&=&
-P_0^UP_{U,0}+a\no
&=&
-2P^+_0P^-_0-P^I_0P_{0,I}-P_0^{D-2}P_0^{D-2}+a\no
&=&
2H-P^I_0P_{0,I}-P_0^{D-2}P_0^{D-2}+a,
\label{mass}
\eqne
where  $a$ is the conventional constant introduced because of ordering ambiguities. 
For the bosonic membrane we have $a=-4$ and $a=-2$ for the closed and open cases, respectively, 
which follows from a generalization 
of the usual string argument \cite{Brink:1986ja}. For the space-time supersymmetric membrane 
one has $a=0$ for both cases. From eqs. (\ref{B}) and (\ref{tilde B}) 
we have $P^{D-2}_0=-{\sqrt{g}}\int d^2\xi B$ 
and $P^{D-2}_0=-{\sqrt{g}}\int d^2\xi \tilde B$ 
for the bosonic and supermembrane, respectively.

For $g=0$ we have that
the mass-shell condition w.r.t. the $(D-1)$-dimensional Poincar\'e group and the 
$D$-dimensional one are equivalent. 
We find 
\eqnb
(m^{(0)})^2\equiv 2H_0-P^I_0P_{0,I}+a=m^2+{\mathcal O}(g)
\eqne
and a discrete mass-spectrum. If we now apply unitary transformations to order $N$ we will get
\eqnb
(m^{(N)})^2\equiv 2H_N-P^I_0P_{0,I}+a=m^2+P^{(N),D-2}_0P^{(N),D-2}_0+{\mathcal O}(g^{N+1}),
\eqne
so that
\eqnb
(m)^2&=&(m^{(N)})^2-P^{(N),D-2}_0P^{(N),D-2}_0+\mathcal O(g^{N+1}).
\label{mass m N}
\eqne
The mass spectrum is, therefore, given by string spectrum corrected by the $(D-2)$-component of the 
momentum. 
Concentrating on the bosonic membrane, we have $B_0^{(N)}=-\frac{1}{\sqrt{g}}P^{(N),D-2}_0$ and we find, 
by eq.\ (\ref{B field}), 
\eqnb
B^{(N)}_0&\equiv&\int d^2\xi B^{(N)}\no
&=&-\sum_{(a)}\left[i n q^{(N)}_{(-a)}\alpha^{(N)}_{0,(a)} \right.\no
&+&
\left.
\sum_{m>0, (a)}\frac{n}{m}\left(\alpha^{(N)}_{-m,(-a)}\alpha^{(N)}_{m,(a)}+
\tilde\alpha^{(N)}_{-m,(-a)}\tilde\alpha^{(N)}_{m,(a)}\right)\right].
\eqne
Note that one could, in principle, add a normal ordering constant to $B_0^{(N)}$. Evaluating it by 
a zeta-function regularization yields the value zero. Evaluating the commutator of this term 
with $q_n^{(N),I}$, $\alpha^{(N),I}_{m,n}$ and $\tilde\alpha^{(N),I}_{m,n}$ 
gives that this operator counts the mode number w.r.t.\ the $\xi^2$-direction,
\eqnb
\left[B^{(N)}_0,q^{(N)}_n\right]&=&nq^{(N)}_n\no
\left[B^{(N)}_0,\alpha^{(N),I}_{m,n}\right]&=&n\alpha^{(N),I}_{m,n}\no
\left[B^{(N)}_0,\tilde\alpha^{(N),I}_{m,n}\right]&=&n\tilde\alpha^{(N),I}_{m,n}.
\eqne
Furthermore, $B^{(N)}_0\left|0,p_0\right>_N=0$. The operators $q_n$ need to be applied in the 
form $\exp(i\sum_n \tilde k_{-n,I}q^I_{n})$ to get an eigenstate of the Hamiltonian. The resulting state is, however, not an 
eigenstate of $B^{(N)}_0$ unless $\tilde k_n=0$, for $n\neq 0$. We restrict ourselves, 
therefore, to $\tilde k_n=0$ for $n\neq 0$.
Using the result in eq.\ (\ref{mass m N}) one will find
\eqnb
(m)^2&=&(m^{(N)})^2-gn^2+\mathcal O(g^{N+1}),
\label{mass at order N}
\eqne
where $n$ is the mode number of the eigenstate in the $\xi^2$-direction.
Thus we see that the degeneracy of the excited states will be lifted and the mass spectrum will be 
split due to the 
mode number in the $\xi^2$-direction. The splitting is exact to all orders and independent of 
the order 
of the perturbation. The groundstate mass is not corrected. As can be seen from eq.\ 
(\ref{mass at order N}), we still have massless string-like states at the first excited level with $n=0$. 
One would expect that when $gn^2\sim 1$ the perturbation theory is not valid. This implies that 
our perturbation theory would be valid for $\left|n\right|<\frac{1}{\sqrt{g}}$ that is, for fixed 
$g$, the magnitude
of the mode number in the $\xi^2$-direction cannot take arbitrarily large values.

For the supermembrane, $B_0^{(N)}$ is replaced by $\tilde B_0^{(N)}$, which again counts the modes 
in the $\xi^2$-direction. Therefore, we get exactly the same result for the mass spectrum, 
i.e.\ the splitting is of the same form as above, where $n$ now also includes fermionic 
excitations. In particular, the 
mass of the groundstate will not get any corrections and we will have massless states. 

Note that for a closed membrane in the $\xi^2$-direction, we will get consistency conditions
on the eigenvalues of $P^{D-2}_0$ due to the gauge choice, eq.\ (\ref{our gauge}). 
This choice is consistent for closed membranes in the $\xi^2$-direction only if $X_0^{D-2}$ is 
periodic. Since $\xi^2$ is periodic with period $2\pi$ we have $X_0^{D-2}$ is 
periodic with period $2\pi/\sqrt g$. Consequently, $P_0^{D-2}$ is quantized and 
take values $n\sqrt{g} $, 
$n\in\mathcal Z$. Thus, from eq.\ (\ref{mass}) one will get
\eqnb
m^2=2H-P^I_0P^I_0-gn^2+a.
\eqne
This is exactly the same spectrum as above. It is a non-trivial check of consistency that these 
two independent derivations are in agreement with each other. It also shows that the magnitude of the 
splitting is non-perturbative.

One may repeat the analysis for an open membrane or semi-open membrane. This is straightforward and the
spectrum will be of the same principal form. In particular, it will contain massless states.

As mentioned 
in section three, we have set $f(\xi^2)=0$, which by consistency, restricts the choices of 
boundary conditions. Let us briefly consider the case when $f$ is non-zero. 
Upon quantization, the function will become an operator $\hat{f}$
depending, in general, on the basic operators $\alpha^I_{n,m}$ and $q^I_n$. In order to 
determine the spectrum 
one needs to
diagonalize $B_0$. The resulting eigenvalues will, therefore, depend on $\int d^2\xi\hat{f}$ and may 
result in
a discrete or continous spectrum. If we, for simplicity, assume that $\int d^2\xi\hat{f}$ is 
diagonal for the same set of
states as with $\hat{f}=0$, then the eigenvalues, $f_0$, of $\int d^2\xi\hat{f}$ will have 
the effect that $B_0$ is changed into $B_0+f_0$.
Therefore, the spectrum in eq. (\ref{mass at order N}) is shifted to
\eqnb
m^2&=&(m^{(N)})^2-g\left(n+2\pi f_0^{(M)}\right)^2,
\eqne
where $(M)$ is a collection of quantum numbers labeling the states.

For closed membranes in the $\xi^2-$direction we still have to fulfill the quantization condition of the momentum,
discussed above.
In this case, the resulting spectrum is simply shifted. 
For the supermembrane, this will imply that there are no massless states, which in turn implies that 
supersymmetry is spontaneously broken. 
For the bosonic case, one 
cannot rule this out. The difference here is that 
massless states for the unperturbed theory are excited states. Then it may be possible to have 
$n+2\pi f_0^{(M)}=0$.

For the open case, the situation may be very different. First of all, the spectrum may either be continous or discrete, 
depending on $\hat f$. Secondly, even for a discrete spectrum, $2\pi f_0^{(M)}$ may not be an integer.
The simplest case is when $\hat{f}=C$ is a constant. The groundstate will have $m^2$ 
that is shifted by a constant value, $-g(2\pi)^4C^2$.
If this value is an integer then the spectrum of the excited states will remain the same. The 
energies of the eigenstates will be permuted within a multiplet.
Massless states are only possible for integer eigenvalues of $2\pi f_0^{(M)}$, and 
only for the bosonic case, as discussed above.

\section{Conclusions}
We have in this paper shown that stretched membrane configurations are essentially described by 
string-like excitations. We have suggested that these string-like modes could play the role of 
elementary membrane states. There are many interesting and important questions that need to be 
answered. First of all the question of consistency of the perturbation theory, e.g.\ the 
convergence, as well as the full $27$- or $11$-dimensional Lorentz symmetry. 

One would also like to probe the non-perturbative properties.  If the conjecture that the 
original membrane action describes a second quantized theory, then it is natural to expect 
that non-perturbative effects will include interaction among the string-like states. We hope to be 
able to come back to this important question in the future.

Finally, it would be interesting and important to further study the matrix model calculations in the
light of our presented results.

\vspace{5mm}

\noindent {\bf Acknowledgement.} We are indebted to Anders Westerberg for helpful
discussions and comments on the manuscript.


\appendix

\sect{Lightcone coordinate conventions}
\eqnb
A^{+}&=&\frac {1}{\sqrt {2}}\left(A^0+A^{D-1}\right)\\
A^{-}&=&\frac {1}{\sqrt {2}}\left(-A^0+A^{D-1}\right)\\
A_UB^U&=&A^+B^- + A^-B^+ + A_IB_I + A_{D-2}B_{D-2},
\eqne
where $U=1,\ldots,D-1$ and $I=1,\ldots,D-3$. We use this explicit basis of the gamma matrices
\eqnb
\Gamma^+&=&\onematrix_{16}\otimes
\left(
\begin{array}{cc}
0&0\\
\sqrt{2} i&0
\end{array}
\right)\\
\Gamma^-&=&\onematrix_{16}\otimes
\left(
\begin{array}{cc}
0&\sqrt{2} i\\
0&0
\end{array}
\right)\\
\Gamma_A&=&\gamma_A\otimes
\left(
\begin{array}{cc}
1&0\\
0&-1
\end{array}
\right).
\eqne
where $A=1,\ldots,D-2$ and $\gamma_9$ can be taken to be
\eqnb
\gamma_9&=&\left(
\begin{array}{cc}
\onematrix_8&0\\
0&-\onematrix_8
\end{array}
\right).
\eqne
The $\bar \theta$ is defined by
\eqnb
\bar\theta&=&\theta^t\Gamma^0,
\eqne
which in our basis is
\eqnb
&&\bar\theta^{2\beta-1}=i\theta^{2\beta}\no
&&\bar\theta^{2\beta}=-i\theta^{2\beta-1}\hspace{1cm} \beta=1,\ldots,16
\eqne

\section{A toy model}
Let us show how our construction works for a toy model that captures many of 
the features of the membrane. In particular, it will expose the non-trivial ordering 
one gets by our construction. The theory is defined by
\eqnb
H&=&H_0+gH_1\no
H_0&=&\frac{1}{2}p^2+a^\dagger a\no
H_1&=&(a^\dagger)^3a+a^\dagger\left(a\right)^3,
\eqne
with the following non-zero Poisson brackets
\eqnb
\left\{q,p\right\}&=&1\no
\left\{a,a^\dagger\right\}&=&-i.
\eqne
Let us construct, order by order, the canonical transformation which maps $H_0$ to $H_1$.
To first order we need to solve
\eqnb
H_{(1)}&\equiv&\exp(-g\ {\ad}_{G_1})H_0\no
&=&
H_0+g\left\{H_0,G_1\right\}+\frac{g^2}{2}\left\{\left\{H_0,G_1\right\},G_1\right\}\no
&=&H+\mathcal O(g^2).
\eqne
One easily finds the following solution
\eqnb
G_1&=&\frac{i}{2}\left[(a^\dagger)^3a-a^\dagger\left(a\right)^3\right].\label{g-one}
\eqne
Defining new coordinates by the canonical transformation, one will get
\eqnb
q_{(1)}&=&\exp(-g\ \ad_{G_1})q=q\no
p_{(1)}&=&\exp(-g\ \ad_{G_1})p=p\no
a_{(1)}&=&\exp(-g\ \ad_{G_1})a=a+\frac{1}{2}g\left(3(a^\dagger)^2a-a^3\right)
+\frac{1}{8}g^2\left(3(a^\dagger)^4a+6(a^\dagger)^2a^3
\right.\no
&+&
\left.
3a^5\right)\no
a^\dagger_{(1)}&=&\exp(-g\ \ad_{G_1})a^\dagger=a^\dagger+\frac{1}{2}g\left(3(a^\dagger)a^2-(a^\dagger)^3\right)
+\frac{1}{8}g^2\left(3a^\dagger a^4+6(a^\dagger)^3a^2
\right.\no
&+&
\left.
3(a^\dagger)^5\right).
\label{a eqn}
\eqne
In terms of the new coordinates, the Hamiltonian  $H_{(1)}$ has the same form as $H_0$. 
Expanding in the old 
coordinates yields
\eqnb
H_{(1)}&=&\frac{1}{2}p_{(1)}^2+a^\dagger_{(1)}a_{(1)}\no
&=&
H(p,a,a^\dagger)+4g^2(a^\dagger)^3a^3+\mathcal O(g^3)\no
&=&
H(p,a,a^\dagger)+4g^2(a_{(1)}^\dagger)^3a_{(1)}^3+\mathcal O(g^3),
\eqne
where we in the last line have used eq.\ (\ref{a eqn}). The $g^2$-term can be transformed away with an
additional canonical transformation
\eqnb
\exp(-g^2\ {\ad}_{G_2})H_{(1)}&=&H(p,a,a^\dagger)+\mathcal O(g^3).
\eqne
The result one finds is
\eqnb
G_2&=&4K_{(1)}(a_{(1)}^\dagger)^3a_{(1)}^3\label{g-two}
\eqne
where
\eqnb
K_{(1)}&=&\frac{1}{2}\left(q_{(1)} p_{(1)}^{-1}+p_{(1)}^{-1}q_{(1)} \right),
\eqne
and satisfies
\eqnb
\left\{H_{(1)},K_{(1)}\right\}&=&-1.
\eqne

We have now constructed the infinitesimal canonical generators that mapsthe unperturbed 
Hamiltonian to the perturbed one to order two in perturbation theory. This we have done classically. 
Let us now follow the procedure outlined in 
section five to construct a well-defined quantum Hamiltonian by applying infinitesimal unitary 
transformations to the unperturbed quantum version of $H_0$. 
The Poisson brackets are exchanged with the commutators 
\eqnb
\left[q,p\right]&=&i\no
\left[a,a^\dagger\right]&=&1.
\eqne
To zeroth order in perturbation theory one can define a groundstate by
\eqnb
a\left|0,p_0\right>_0&=&0\no
p\left|0,p_0\right>_0&=&p_0\left|0,p_0\right>_0.
\eqne
From this groundstate one can now construct the states of this theory
\eqnb
\left|N,p_0\right>&=&\frac{1}{\sqrt{N!}}\left(a^\dagger\right)^{N}\left|0,p_0\right>\no
\left|0,p_0+k_0\right>&=&\exp(ik_0q)\left|0,p_0\right>,
\eqne
where $k_0$ is a constant. Let us make a unitary transformation, up to order two in $g$, 
constructed from $G_1$ in eq. (\ref{g-one}), 
\eqnb
U_1\equiv \exp\left(ig:_0G_1:_0\right)&=&1+ig:_0G_1:_0-
\frac{1}{2}g^2\left(:_0G_1:_0\right)^2+\mathcal O(g^3).
\eqne
We use this transformation to define a new groundstate as
\eqnb
\left|0,p_0\right>_1&=&\exp\left(ig:_0G_1:_0\right)\left|0,p_0\right>_0\no
&=&\left(1-\frac{g}{2}\left[(a^\dagger)^3a-a^\dagger\left(a\right)^3\right]+
\frac{g^2}{8}\left[(a^\dagger)^3a-a^\dagger\left(a\right)^3\right]\right.\no
&\times &
\left.\left[(a^\dagger)^3a-a^\dagger\left(a\right)^3\right]\right)\left|0,p_0\right>_0+\mathcal O(g^3)
=\left|0,p_0\right>_0.
\eqne
Therefore, the first order groundstate 
is the same as the zeroth order one\footnote{This is not true for more general models, 
for example, where
$H_1$  contains terms of the form $(a^\dagger)^4+a^4$.}. Let us also transform the operators 
\eqnb
q_{(1)}&\equiv &U_1qU_1^\dagger=q\no
p_{(1)}&\equiv &U_1pU_1^\dagger=p\no
a_{(1)}&\equiv& U_1aU_1^\dagger=a+\frac{1}{2}g\left(3(a^\dagger)^2a-a^3\right)
+\frac{1}{8}g^2\left(3(a^\dagger)^4a+3(a^\dagger)^2a^3
\right.
\no
&+&
6a^\dagger a^2a^\dagger a-\left.3a^2(a^\dagger)^2a+3a^5\right)\no
a^\dagger_{(1)}&\equiv& U_1a^\dagger U_1^\dagger=a^\dagger+\frac{1}{2}g\left(3a^\dagger a^2-(a^\dagger)^3\right)
+\frac{1}{8}g^2\left(3a^\dagger a^4+3(a^\dagger)^3a^2
\right.
\no
&+&
\left.
6a^\dagger a(a^\dagger)^2a-3a^\dagger a^2(a^\dagger)^2+3(a^\dagger)^5\right).
\label{a eqn quantum}
\eqne
Inserting this, one gets
\eqnb
H_{(1)}&=&\frac{1}{2}p_{(1)}^2+a_{(1)}^\dagger a_{(1)}\no
&=&\frac{1}{2}p^2+a^\dagger a+g\left((a^\dagger)^3a+a^\dagger a^3\right)
+g^2\left((a^\dagger)^3a^3+\frac{3}{2}a^\dagger a^2(a^\dagger)^2a
\right.
\no
&+&
\left.
\frac{3}{4}(a^\dagger)^2a^2a^\dagger a
+\frac{3}{4}a^\dagger a(a^\dagger)^2a^2
\right)+\mathcal O(g^3)\no
&=&
H+g^2\left((a_{(1)}^\dagger)^3a_{(1)}^3
+\frac{3}{2}a_{(1)}^\dagger a_{(1)}^2(a_{(1)}^\dagger)^2a_{(1)}
+\frac{3}{4}(a_{(1)}^\dagger)^2a_{(1)}^2a_{(1)}^\dagger a_{(1)}
\right.\no
&+&
\left.
\frac{3}{4}a_{(1)}^\dagger a_{(1)}(a_{(1)}^\dagger)^2a_{(1)}^2\right)
+
\mathcal O(g^3).
\eqne
Thus, to first order, the quantum Hamiltonians 
$H$ and $H_{(1)}$ coincide.

Let us now use $G_2$, given by eq. (\ref{g-two}), to 
construct the next unitary transformation,
\eqnb
U_2&=&\exp\left(ig^2:_1G_2:_1\right).
\eqne
We then define the Hamiltonian 
\eqnb
H_{(2)}&\equiv&\frac{1}{2}p_{(2)}^2+a_{(2)}^\dagger a_{(2)}=U_2H_{(1)}U_2^\dagger\no
&=&
H+g^2\left((a_{(1)}^\dagger)^3a_{(1)}^3
+\frac{3}{2}a_{(1)}^\dagger a_{(1)}^2(a_{(1)}^\dagger)^2a_{(1)}
+\frac{3}{4}(a_{(1)}^\dagger)^2a_{(1)}^2 a_{(1)}^\dagger a_{(1)}
\right.
\no
&+&
\left.
\frac{3}{4}a_{(1)}^\dagger a_{(1)}(a_{(1)}^\dagger)^2a_{(1)}^2 
-4(a_{(1)}^\dagger)^3a_{(1)}^3
\right)+\mathcal O(g^3).
\eqne
As can be seen from this equation, the Hamiltonian $H^{(2)}$ differs from $H$, up to 
order two, only 
by ordering terms. The groundstate to this order, $\left|0,p_0\right>_{2}=U_2\left|0,p_0\right>_{1}$, 
is the same as the zeroth order vacuum. The operators are transformed to
\eqnb
q_{(2)}&=&U_2q_{(1)}U_2^\dagger =q_{(1)}+2g^2\left(q_{(1)}p^{-2}_{(1)}+p^{-2}_{(1)}q_{(1)}\right)(a_{(1)}^\dagger)^3 (a_{(1)})^3\no
p_{(2)}&=&U_2 p_{(1)}U_2^\dagger =p_{(1)}+4g^2p^{-1}_{(1)}(a_{(1)}^\dagger)^3 (a_{(1)})^3\no
a_{(2)}&=&U_2 a_{(1)}U_2^\dagger =a_{(1)}-12g^2iK_{(1)}(a_{(1)}^\dagger)^3 (a_{(1)})^2\no
a_{(2)}^\dagger&=&U_2 a_{(1)}^\dagger U_2^\dagger =a_{(1)}+12g^2iK_{(1)}(a_{(1)}^\dagger)^2 (a_{(1)})^3,
\eqne
where $q_{(1)}$, $p_{(1)}$, $a_{(1)}$ and $a_{(1)}^\dagger$ are given by eq.\ (\ref{a eqn quantum}). 
As can be seen from the equations, even if the theory is simple and we only consider the 
transformations up to order two, the resulting expressions are relatively complicated.

\end{document}